\documentclass[prl,aps,twocolumn,floatfix]{revtex4}
\usepackage{amsmath}
\usepackage{amssymb}
\usepackage{graphicx}
\newcommand{\be}{\begin{equation}}
\newcommand{\ee}{\end{equation}}
\newcommand{\chill}{$\chi$LL }

\begin{document}
\title{Temperature dependence of the tunneling amplitude between
   Quantum Hall edges}
\author{Roberto D'Agosta}
\email{dagostar@missouri.edu}
\affiliation{NEST-INFM}
\affiliation{Department of Physics and Astronomy, University of
   Missouri - Columbia, 65211, Columbia, Missouri, USA}
  \author{Giovanni Vignale}
\affiliation{NEST-INFM}
\affiliation{Department of Physics and Astronomy, University of
   Missouri - Columbia, 65211, Columbia, Missouri, USA}
\author{Roberto Raimondi}
\affiliation{NEST-INFM}
\affiliation{Dipartimento di Fisica, Universit\`a di Roma Tre,
  via della Vasca Navale 84, 00146, Roma, Italy}
  \begin{abstract}
Recent experiments have studied the  tunneling current  between the
edges of  a fractional quantum Hall liquid as a function of 
temperature and voltage. The results of the experiment are puzzling 
because at ``high" temperature  ($600-900$ mK)  the behavior of the 
tunneling conductance is consistent with the theory of tunneling 
between chiral Luttinger liquids, but at low temperature it strongly 
deviates from that prediction dropping to zero with decreasing 
temperature. In this paper we suggest a possible explanation of this 
behavior in terms of the strong temperature dependence of the 
tunneling amplitude.
  \end{abstract}
\date{\today}
%\showpacs{}
\maketitle

%In the last twenty years the quantum Hall effect has been a 
%rich source of fascinating phenomena and new theoretical concepts.
In the last twenty years quantum Hall systems have been a rich source
of information about the physics of correlated electron systems.
%One  interesting problem  is the subtle connection
%between the integer and the fractional effects.
One example is the edge of a Fractional quantum Hall
system which represents one of the best realization of a strictly one
dimensional interacting system.
%  Luttinger liquid. 
%This occurs for example when describing the role of the edges
%in Fractional quantum Hall systems. 
%In fact, whereas in integer quantum Hall systems,
%the edges are described in terms of one-dimensional Fermi liquids,
Indeed, Wen showed that the low-energy density excitations localized 
along the edges of a fractional quantum Hall liquid  are 
effectively described by a chiral  Luttinger Liquid
($\chi$LL) \cite{Wen1990,Wen1991a}, with the effective interaction
parameter given by the {\sl bulk} filling factor $\nu_b$. 
%
%In particular the fractional quantum 
%Hall effect (FQHE) is effectively due to the electron-electron 
%interaction, while the strong magnetic field forces the electrons to 
%occupy the lowest Landau level.
%Whereas the main features of the integer quantum Hall effect
%are captured by considering the interplay of magnetic field and 
%disorder, the understanding of the fractional quantum 
%Hall effect (FQHE) requires to take into account
%the correlation amongst the electrons.  
%An example is provided by the pioneering work of
%In a subsequent development 
%Wen which showed that 
%the low-energy density excitations localized along the edges of a 
%fractional quantum Hall liquid  are effectively described by 
%form what is known as 
%a chiral  Luttinger Liquid
%($\chi$LL) \cite{Wen1990,Wen1991a} -- a model 1D system in which the 
%correlation functions exhibit  power law behaviors with exponent 
%determined by the bulk filling factor $\nu_b$\cite{Wen1991a,Kane1992}.
%We now believe that the  $\chi$LL  behavior is generic to the edge 
%density fluctuations of a 2D electron liquid in the lowest Landau 
%level, irrespective of whether  the system exhibits the quantum Hall 
%effect or not\cite{DAgosta2003}.

Tunneling experiments offer an effective way to probe in detail the 
predictions of the \chill model \cite{Wen1991a,Kane1992,DAgosta2003}.  
In particular, measurements of the 
tunneling current from an external metallic  gate into the edge of a 
2DEG have beautifully confirmed the theoretical prediction
of a tunneling current proportional to $V_T^{\frac{1}{\nu_b}}$ for 
$k_BT \ll eV_T$ ($V_T$ being the potential difference between the edge 
and the gate)
% --  a behavior that can be qualitatively understood in 
%terms of the ``orthogonality catastrophe" 
%concept
\cite{Milliken1995,Chang1996,Grayson1998}.
The tunneling of fractionally charged quasiparticles between the 
edges of a fractional quantum Hall liquid has also been studied 
experimentally by several groups~\cite{Chung2003,Chung2003a,Roddaro2002}. 
Of particular interest to us are the recent measurements performed in 
the weak tunneling regime at $\nu_b=1/3$ by Roddaro 
{\it et al.}~\cite{Roddaro2002}. 
According to the theory, one expects that,
in this experiment, the  tunneling current  must  scale as 
$V_T^{2\nu_b -2}$ for $k_BT\ll eV_T$ and  be linear in $V_T$ for 
$V_T\ll k_BT$.  The zero-bias tunneling conductance,  
$\left.\frac{dI_T}{dV_T}\right\vert_{V_T=0}$, furthermore, should  
grow as $T^{2\nu_b-2}$ with decreasing 
temperature~\cite{Wen1991a,Kane1992,DAgosta2003}. 
Contrary to this expectation, 
%What was actually seen is quite different.  
while in the temperature range $600$ mK $<T<900$ 
mK one observes   
%there is just 
%a hint of 
a growing conductance with decreasing 
temperature,
%.  As the temperature drops 
below $600$ mK one 
sees a dramatic drop in the tunneling conductance.
 % rapidly approaching zero. 
We emphasize that this is in glaring contrast not only with the prediction 
of the weak tunneling theory, but also with  the 
exact theory~\cite{Fendley1995} valid in both weak- and strong-tunneling regimes. 
%which predicts a stronger rate of 
%increase of the conductance with decreasing temperature.   

In this Letter we argue that these puzzling data may be explained by a strong 
temperature dependence of the inter-edge tunneling amplitude.   More 
precisely, we will show that the spatial separation between the edges 
of a fractional quantum Hall liquid increases with decreasing 
temperature, resulting in a rapid loss of overlap between the edges 
and a consequent collapse of the tunneling amplitude on a temperature 
scale $T_0$ quite comparable to the $600$ mK observed in the 
experiment.

\begin{figure}[t!]
\includegraphics[width=6cm,clip]{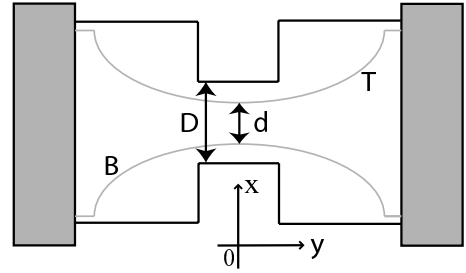}
\caption{Simple scheme of the experimental setup. The current is carried by the
quasi-particle in the edge states that are forced to stay close by the presence
of the geometrical constriction of width $D$. Note that, inside the
constriction, the edges are at the distance $d$  with $d\ll D$. We assumed as
the $0$ for the $y$ coordinate the position where the edge distance is minimal
and the tunneling takes place (see Ref.
\onlinecite{DAgosta2003} for further details on the choice of the reference
frame and Ref. \onlinecite{Roddaro2002} for the details about the actual
device).}
\label{simpledevice}
\end{figure}
The common starting point for calculating the differential tunneling 
conductance is a model consisting of two \chill s (the two edges) 
coupled  by the tunneling hamiltonian
\be
\label{tunnelinghamiltonian}
H_T=
\Gamma \hat \Psi^\dagger_T(0)\hat \Psi_B(0)+\Gamma^*\hat 
\Psi^\dagger_B(0)\hat \Psi_T(0)
\ee
where
$\Gamma$ is a phenomenological tunneling amplitude and the operators 
$\hat \Psi_{T(B)}(0)$ destroy a quasiparticle of fractional charge 
$e^*=\nu_be$  at a point ``$0$"  in
the top or bottom edge respectively (see Fig. \ref{simpledevice}). 
A standard perturbative calculation leads to the following expression 
for the differential tunneling  conductance $G=\frac{dI_T}{dV_T}$ at 
$V_T=0$\cite{Wen1991a,DAgosta2003}:
\be\label{tunnelingcurrent}
G=\frac{e^2}{h} 
\left(\frac{\Gamma}{\hbar v}\right)^2 
\left(\frac{k_BT}{\hbar\omega_0}\right)^{2\nu_b-2} B(\nu_b,\nu_b)
\ee
where $v$ is the velocity of the edge modes, $\omega_0$ is an ultraviolet frequency cutoff related to the microscopic cutoff length $a$ by 
$\omega_0=\frac{v}{a}$,  and $B(x,y)$ is the Euler beta function 
\cite{Abramowitz1964}. 

\begin{figure}[t!]
\includegraphics[width=8cm,clip]{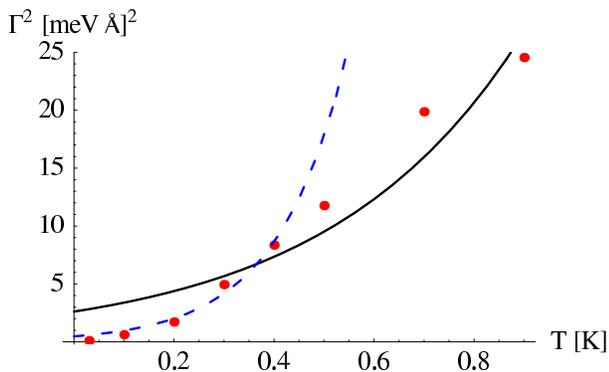}
\caption{The variation of $|\Gamma|^2$ with temperature. The points (red) are
the experimental results \cite{Roddaro2002} obtained from the the evaluation of
Eq. (\ref{tunnelingcurrent}). The lines are two fits with the function $g
\exp(T/T_0)$. The solid line (black) is a fit with all the experimental data and
gives $g=2.6~(\mathrm{meV\AA})^2$ and $T_0=400$ mK while for the dashed line
(blue) we have considered only temperatures 
below $400$ mK and gives the estimates
$g=0.5~(\mathrm{meV\AA})^2$ and $T_0=140$ mK.}
\label{experimental}
\end{figure}

It is normally assumed that the tunneling 
amplitude is independent of temperature: if this were true it would 
imply $G\propto T^{2 \nu_b-2}$, increasing with decreasing 
temperature.   However, an analysis of the experimental data  of 
Ref.~\onlinecite{Roddaro2002}, shows that the situation is quite 
different.  We extract the value of $\Gamma$ from the measured values 
of the conductance simply by inverting  Eq.~(\ref{tunnelingcurrent}), 
using $v\simeq4\times10^5~\mathrm{m/s}$ for the edge wave 
velocity~\cite{Aleiner1994} and
$a=100~\mathrm{\AA}$ \cite{Wen1991a} for the ultraviolet length 
cutoff.  The values of $\Gamma$ obtained in this manner are shown as 
solid dots in Fig.~\ref{experimental}. 
Notice that $\Gamma$ increases rapidly with temperature 
below about $600$~mK and more slowly for 
$T>600$~mK.

%How to understand this unexpected behavior?    
To understand this unexpected behavior, we begin by recalling that the
%Recall that the 
tunneling amplitude arises from the overlap of single particle states 
localized in front of each other on the top and bottom edges.  For two 
coherent states in the lowest Landau level centered respectively at 
$0,T$ and $0,B$ (see Fig.~\ref{simpledevice})  the matrix element of the 
noninteracting hamiltonian is (up to an irrelevant phase factor) 
\be 
\label{GammaFormula}
\Gamma = \frac{\hbar^2}{2 m^* \ell}e^{-\frac{d^2}{4\ell^2}}
\ee 
where $d$ is the distance between the edges at the center of the 
constriction, $\ell$ is the magnetic length, and $m^*$ is the 
effective mass.    It is important to realize that $d$ is typically 
much smaller than the geometric separation, $D$, between the split 
gates   (in the experiments of Ref.~\cite{Roddaro2002}, with $m^* 
\simeq 0.067 m$ for GaAs, and $\ell \simeq 100~\mathrm{\AA}$, one has $d \sim 
3-5 \ell$ \footnote{This estimate follows from Eq. (\ref{GammaFormula}) after
the observation that $\Gamma\simeq 5-25~\mathrm{meV \AA}$.},
while $D\sim 30 \ell$)  and that its value is determined 
by equilibrium considerations discussed in detail below.  Due to the 
exponential dependence of $\Gamma$ on $d$ even a relatively small 
variation of $d$ with temperature can have a large effect on 
$\Gamma$.  Moreover we will show that, at low temperatures, $d$ varies linearly
with the temperature.

\begin{figure}[t!]
\includegraphics[width=8cm,clip]{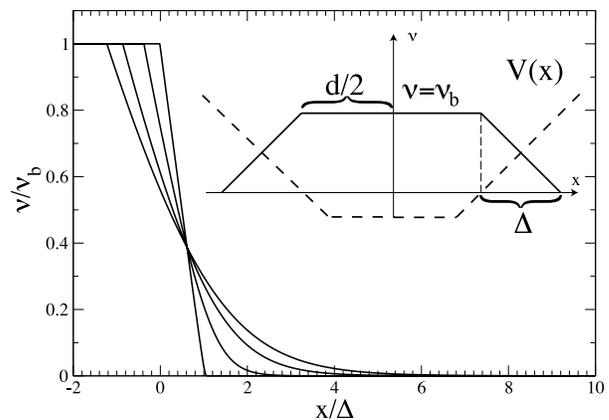}
\caption{The solution of Eq. (\ref{stationary}) for various
temperatures $k_B T/U=0.01,~0.51,~0.81,~1.01$). Inset: Plot of
of the local filling factor profile at $T=0$ (solid line) 
and of the confining potential (dashed line).}
\label{solutionplot}
\end{figure}
Our picture of the  system in shown in the inset of Fig.~\ref{solutionplot}. 
The center of the Hall bar 
is occupied by an incompressible quantum Hall strip of width $d$, 
sandwiched between two compressible  regions of smoothly varying 
density.  Since the density tapers off from the uniform value in the 
incompressible strip  to zero over a distance of several magnetic 
lengths, what we are showing here is essentially the situation 
depicted by Chklovskii {\it et al.} in their classical electrostatic theory 
of  edge channels \cite{Chklovskii1992,Chklovskii1993}.  The density 
profile is determined, at $T=0$,  by minimizing the sum of the 
electrostatic energy and the confinement energy, subject to the 
constraint of having an incompressible strip at the center of the 
system \footnote{In Ref. \onlinecite{Lier1994} and \onlinecite{Oh1997} the
problem of the incompressible strip formation in the 
Integer Quantum Hall effect
has been considered in a self consistent approach. 
%In those articles however
%the assumption of incompressible bulk is not present and the result
%presented cannot be extended to the case of the FQHE.
}. 
In order to arrive at an analytically tractable model we 
assume that the system is translationally invariant in the 
$y$ direction  (i.e.,  the density profile depends only on $x$)  and 
that the electron-electron interaction is  screened, due to the 
presence of the split gates,  beyond a characteristic screening 
length $\lambda$, also of the order of several magnetic lengths. 
We also assume that the system is symmetric with respect to
$x=0$ and study below only the part with $x>0$:  thus we neglect any 
interaction between the top and the bottom part of the system. None 
of these simplifications alters the qualitative features of the 
solution.   

The total energy  associated with a given  density 
profile $n(x)$ can be written as
\be
E=\frac{\pi e^2 \lambda L}{\epsilon_b}\int n(x)^2~ dx+L\int V(x) n(x)~dx
\ee
where $\epsilon_b$  is
the dielectric constant, $V(x)$ is 
the external confining potential (from gates, etc.),  and $L$  is the 
length of the system in the $y$ direction. The integral runs over the 
top inhomogeneous region. At finite temperature, we also need to 
include the electronic entropy.  This is obtained in the standard way 
from the assumption that the local filling factor $\nu(x) \equiv 2 
\pi \ell^2 n(x)$  gives the probability of a single particle state 
centered at $x$ in the lowest Landau level to be occupied.  Thus, we 
have
\be\label{entropy}
\begin{split}
S=-\frac{k_B L}{2\pi \ell^2}\int &\left\{\nu(x)\ln\nu(x)\right.\\
&\left.+[1-\nu(x)]\ln[1-\nu(x)]\right\}~dx.
\end{split}
\ee

The edge density profile  is now computed  from the requirement that 
the free energy  $F=E-TS$ is stationary with respect to small
variations of the density, subject to the constraint of global 
particle number conservation and with the further condition 
$\nu(x)=\nu_b$ at the edge of the incompressible strip (notice that 
the position of this  edge is itself to be determined).   These 
requirements easily lead to the equation
\be
U \nu(x) + V(x)+k_B T
\ln\left[\frac{\nu(x)}{1-\nu(x)}\right]=\mu
\label{stationary}
\ee
which must be satisfied in the compressible region determined by  the
conditions $0<\nu(x)<\nu_b$. Here $U=\lambda
  e^2/\epsilon_b \ell^2$ represents a typical interaction energy,  
$\mu$ is the chemical potential, which fixes the total particle 
number, and the edge of the incompressible strip occurs at the 
position for which $\nu(x=d(T)/2)=\nu_b$ (cf. Fig. \ref{solutionplot}). 
For the sake of
simplicity we take the position of the edge at $T=0$ as the origin of the
coordinate, $x\to x-d(0)/2$.
%To fix the ideas we choose this position
%as the origin so that $\nu(0)=\nu_b$.
To proceed,  we assume that around this point 
the external potential can be linearly expanded,
 $V(x)=e{\cal E} x$\footnote{The constant term $V(0)$ represents
  merely a shift in the chemical potential.}, where ${\cal E}$  is 
the electric field. 
Eq. (\ref{stationary}) admits an elegant solution in this case.
However, we expect that non-linear terms yield no qualitative differences
as long as one considers not too high temperatures.
To 
begin with, by setting $T=0$,  we easily find the zero-temperature 
solution
\be
\nu_0(x) =\left\{ 
\begin{array}{lc}
\nu_b \left( 1 - 
\frac{x}{\Delta} \right), & 0<x<\Delta\\
\nu_b, & x<0\\
0, & x>\Delta
\end{array}\right.
\ee
where 
$\Delta =\frac{U}{e{\cal E}}\nu_b$ 
is the 
width of the compressible region %and we have chosen
and $\mu=U \nu_b$.
% so that 
%the edge of the incompressible strip occurs precisely at $x=0$. 

At 
finite temperature, the chemical potential must be chosen in such a 
way that the total particle number remains the same as at $T=0$: 
therefore we must have
\be 
\label{numberconservation}
\int_{x_0}^\infty \nu(x)~dx = 
\int_{x_0}^\Delta \nu_0(x)~dx  = -\nu_b x_0 
+\frac{\Delta\nu_b}{2},
\ee
where the position of the edge, 
$x_0$,  is determined by the condition  $\nu(x_0)=\nu_b$.  
Due to the linearity of the external potential, 
the integral on the left hand side  of Eq.~(\ref{numberconservation}) 
can be evaluated analytically by a change of variable from 
$x$ to $\nu$ after an integration by parts. This yields
% with the following result:
\be\label{solutionx0}
\begin{split}
\int_{x_0}^\infty 
\nu(x)~dx=&-\nu_bx_0-\frac{\Delta\nu_b}{2}
+\frac{1}{e\mathcal{E}}\left\{
  \mu\nu_b\right.\\
&\left.-k_B T[\nu_b \ln\nu_b 
+(1-\nu_b)\ln(1-\nu_b)]\right\}.
\end{split}
%\Delta\nu_b \left[\frac{1}{2}- \frac{k_B T}{U}\frac{\ln 
%(1-\nu_b)}{\nu_b^2}\right].
\ee
By comparing Eq. (\ref{numberconservation}) and
Eq. (\ref{solutionx0}) we get the temperature shift of the chemical potential
\be\label{mushift}
\mu(T)-\mu(0)=\frac{k_BT}{\nu_b}\left[\nu_b \ln\nu_b+(1-\nu_b)\ln
(1-\nu_b)\right]
\ee
and by evaluating  Eq. (\ref{stationary}) for $x=x_0$, we have
\be\label{edgeposition}
x_0=\frac{k_B T}{U}\frac{\ln 
(1-\nu_b)}{\nu_b^2}\Delta
%\frac{k_B T}{e
%  \mathcal{E}}\frac{\ln(1-\nu_b)}{\nu_b}
\ee
 which yields the effective edge separation by recalling that
$d(T)=d(0)+2x_0$.
Fig.~\ref{solutionplot} shows the numerical solution of 
Eq.~(\ref{stationary})  for $\nu(x)$ obtained for different 
temperatures.  Notice that the edge of the incompressible strip 
shifts inward as predicted by Eq.~(\ref{edgeposition}). 
% As a 
%curiosity, one may show that all the solutions intersect at $\bar \nu 
%= \frac{\nu_b}{\nu_b+(1-\nu_b)^{1-1/\nu_b}}$, $\bar x = 
%\Delta(1-\bar\nu/\nu_b)$.

%\begin{figure}[t!]
%\includegraphics[width=8cm,clip]{solution3nb.eps}
%\caption{The solution of Eq. (\ref{stationary}) for various
%temperatures $k_B T/U=0.01,~0.51,~0.81,~1.01$). Inset: Plot of
%of the local filling factor profile at $T=0$ (solid line) 
%and of the confining potential (dashed line).}
%\label{solutionplot}
%\end{figure}

Putting Eq.~(\ref{edgeposition}) in Eq.~(\ref{GammaFormula}) we finally arrive
at
\be
|\Gamma(T)|^2=|\Gamma(0)|^2 e^{T/T_0}
\ee
where \be
k_B T_0=\left|\frac{\nu_b}{\ln(1-\nu_b)}\right|\frac{\lambda}{2\Delta}
\left(\frac{\nu_b e^2}{\epsilon_bd}\right)~.
%=\left|\frac{\nu_b}{\ln(1-\nu_b)}\right|\frac{e\mathcal{E}\ell^2}{2d}.
\label{tnot}
\ee
From the experiments \cite{Roddaro2002} we estimate that 
$\Delta\simeq 3 \lambda $ \footnote{We estimate that the
screening length $\lambda$ must be smaller than or of the same order of
magnitude of the distance of the electron gas from the surface where 
the gates were etched. In the experiment reported in 
\cite{Roddaro2002} this distance is
$100~\mathrm{nm}\simeq 10\ell$ while $\Delta$ is half the distance between the
split gates which is $\sim 300~\mathrm{nm}$.}
and $d(0)\simeq 4 \ell$: 
thus we obtain $T_0\simeq 600~\mathrm{mK}$ which is comparable with
the value obtained from the fits shown in Fig.~\ref{experimental}.
Since $\Delta=U\nu_b / e{\mathcal E}$, one expects that the characteristic
temperature scale $T_0$ increases by making the confining potential steeper.
This
prediction appears to be qualitatively in agreement with recent
experiments \cite{Roddaro2004a} where the behavior of the tunneling
conductance has been investigated as a function of the gate voltage
controlling the quantum point contact.
For sufficiently negative gate voltage the tunneling conductance
is consistent with the prediction of the \chill
model with a constant $\Gamma$. This in turn is consistent with a 
large characteristic temperature scale $T_0$ as predicted by Eq. (\ref{tnot}).
%such that the
%tunneling amplitude is constant in the experimental temperature range.  

We believe that our electrostatic model, in spite of its simplicity,  
captures the essential aspects of the observed temperature dependence of the 
tunneling amplitude.  The main effect of the temperature 
is to remove particles 
from the incompressible strip transferring them into the zone
that was depleted at $T=0$.  This causes a linear increase in entropy, 
coming primarily from the population of states that were initially 
empty. Let us emphasize that Eq. (\ref{entropy}) takes into account only the
entropy of the compressible strip and that the electrons in the 
incompressible strip are locked in a collective state of essentially 
zero entropy for temperatures below the fractional quantum Hall gap.
%In this point of view at the edge $x_0$ two different phases with different
%entropy coexist.
%:this effect is  only likely to become larger in a more 
%accurate model (e.g., including the fact that the electrons in the 
%incompressible strip are locked in a collective state of essentially 
%zero entropy for temperatures below the fractional quantum Hall gap). 
%Since the $T=0$ state is a state of minimum energy we expect that at 
%finite low temperature the change in
%energy will be quadratic in the edge position while the
%entropic contribution will be linear. Hence, the latter will always 
%win and cause a linear-in-$T$ shift in the position of the edge \footnote{Let us
%point out again that only the quasi-particles near the edge $x=x_0$ can
%contribute to the edge displacement due to the entropic term. 
%Indeed near $x=x_0$ the electrostatic energy compensates for the gap making
%possible the rearrangement of those quasi-particles while well inside the
%bulk the particles are locked by the presence of the gap.}.
Thus we do not expect that the general scenario presented here will be
significantly affected by introducing more realistic features in the 
calculation of the energy and of the confining potential. 

On the other hand, our analysis of the experiment assumes the 
validity of  Eq.~(\ref{tunnelingcurrent}), itself a consequence of 
the weak tunneling theory of Wen.  Recently, there have been 
suggestions  that Eq.~(\ref{tunnelingcurrent}) 
might be invalidated by additional interactions between electrons on 
the {\it same edge},  since these interactions appear to change the 
scaling dimension of the tunneling  \cite{Papa2004}.  
%Should this turn out to be the 
%case the whole analysis of the data based on 
%Eq.~(\ref{tunnelingcurrent}) would have to be reconsidered. However an even
%stronger temperature dependence of the tunneling amplitude should be expected
%in order to explain the increase of the tunneling conductance at zero bias for
%temperature above $600$ mK.
For that mechanism to be effective the long range intra-edge
interaction must be stronger than the inter-edge interaction:
this condition is unlikely to be satisfied in
the present experimental setup.

As a final point, we note that the dependence of the inter-edge 
separation on temperature is not expected to translate into a 
dependence of this quantity on the applied voltage.  Indeed, in the 
present experiment 
this voltage is just the Hall voltage created by the dc current 
injected in the Hall bar~\cite{Roddaro2002}. 
The effect of this 
current is to create different
quasi-particle populations on the two edges. However, in our model, 
this will  cause a rigid shift of both edges in the same direction 
thus leaving the distance between them and hence the tunneling
amplitude unaffected.

In conclusion,  in this Letter we have addressed the problem of determining the
temperature dependence of the tunneling amplitude in the tunneling process
between the edges of a fractional quantum Hall liquid. We have shown
that the temperature modifies in a non trivial way the equilibrium 
distance between the edges, and  therefore the tunneling
amplitude which is a very sensitive function of the temperature.

\begin{acknowledgments}
We are grateful to S. Roddaro, V. Pellegrini, and F. Beltram for useful
discussions and the use of their experimental data. We kindly acknowledge the
hospitality of the Max Planck Institute for the Physics of Complex
Systems in Dresden where part of this work was completed. 
This research was supported
by NEST-INFM PRA-Mesodyf and NSF DMR-0313681. R.D'A. acknowledges the
financial support by NEST-INFM PRA-Mesodyf. 
\end{acknowledgments}

\bibliography{qhe-biblio}
\end{document}